\documentclass[aps,prc,superscriptaddress,twocolumn,nofootinbib]{revtex4}
\usepackage{amsmath,graphicx}
\usepackage[T1]{fontenc}
\usepackage{epsfig}
\usepackage{amssymb}
\usepackage{mathrsfs}
\usepackage{mathcomp}

 \newcommand{\as}{\bar{\alpha}_s}

\def\pmb#1{{\mbox{\boldmath$#1$}}}
\begin{document}
\title{Exclusive final states in the Dipole Model\footnote{Work supported in part by the Marie
    Curie RTN ``MCnet'' (contract number MRTN-CT-2006-035606).}}
\author{Christoffer Flensburg\protect\footnote{\ \ In collaboration with G\"osta Gustafson and Leif L\"onnblad}}
\affiliation{Dept.~of Theoretical Physics,
  S\"olvegatan 14A, S-223 62  Lund, Sweden}
\begin{abstract}
\vspace {-2.45cm}
\begin{flushright}
LU-TP 10-23
\end{flushright}
\vspace {1.6cm}
We have developed a BFKL-based dipole model in impact parameter space, which has successfully predicted a wide range of total cross sections. The model has now been extended to a full event generator which is showing promising results. Some of the difficulties in non-leading order BFKL are discussed, and preliminary results are compared to CDF and ALICE data.
\end{abstract}

\maketitle

\section{Introduction}
A BFKL-based initial state dipole model in impact parameter space has been developed in a series of papers \cite{Avsar:2005iz,Avsar:2006jy,Avsar:2007xg,Flensburg:2008ag,Flensburg:2010kq}. In the Good--Walker approach \cite{Good:1960ba} the two incoming hadronic particles are seen as linear superpositions of eigenstates to the interaction. Miettinen and Pumplin suggested that the interaction eigenstates are parton cascades \cite{Miettinen:1978jb}, which in this model are generated with a dipole model. Only inclusive, and some semi-inclusive, observables have been studied in previous publications, but we now want to use the information in the initial state cascades to produce exclusive final states. Some of the partons in the initial state cascade will be selected as real partons, while others will be seen as virtual fluctuations that will be absorbed. The real partons will be further evolved through final state radiation with ARIADNE \cite{Lonnblad:1992tz} and hadronisation with PYTHIA \cite{Sjostrand:2006za,Andersson:1983ia}.

The way the real partons are selected is not straight forward, and there are many details in higher orders that have large impact on observables. Some of these details may be solved by further study, but others seem to be out of reach from current perturbative calculations. It should be well noted that the results presented here are not the only possible way of extending the inclusive formalism. This short summary will not go into detail on these complications, leaving them for a publication in the near future.

\section{The Cascade}
The cascades are generated from an initial valence state of dipoles in impact parameter space, and is then evolved in rapidity with a BFKL-based splitting probability. The foundation of the cascade is the leading logarithm formalism first used by Mueller and coworkers \cite{Mueller:1993rr, Mueller:1994jq, Mueller:1994gb}:

\begin{eqnarray}
\frac{d\mathcal{P}}{dY}=\frac{\bar{\alpha}}{2\pi}d^2\pmb{z}
\frac{(\pmb{x}-\pmb{y})^2}{(\pmb{x}-\pmb{z})^2 (\pmb{z}-\pmb{y})^2},
\,\,\,\,\,\,\, \mathrm{with}\,\,\, \bar{\alpha} = \frac{3\alpha_s}{\pi}.
\label{eq:dipkernel1}
\end{eqnarray}
Here $\mathcal{P}$ is the emission probability density, integrated over the transverse position $\pmb{z}$ of the emitted gluon. $\pmb{x}$ and $\pmb{y}$ are the transverse positions of the partons in the emitting dipole.

The cascade in our model introduces a number of corrections to Muellers original formulation.
\begin{itemize}
\item \emph{The momenta} of the partons are necessary to determine for many of the following corrections. In DIPSY, each emission gets a transverse recoil from each parent of $\pmb{r}_{t}/\pmb{r}_{t}^2$ where $\pmb{r}_t$ is the transverse distance between the emission and the parent. With the transverse momentum set and the rapidity known from the emission, the $p_+$ of an emission can be calculated through

$$p_+ = p_Te^{-y}$$

where $y$ is the rapidity. Notice that the rapidity is negative for particles incoming along the negative $z$-axis. The negative lightcone momentum depends on the structure of the rest of the event, but will in the cascade be assumed to be that of an on-shell gluon:

$$p_- = p_T e^{y}.$$

To conserve momentum, recoils in $p_T$ and $p_+$ are given to the partons in the emitting dipole. The $p_-$ will be delivered from the colliding partons from the other side, and the limitations are taken into account when the interactions are decided.
\item \emph{Next to leading logarithm effects} are known to be large in BFKL. While this is not a full NLL calculation, it includes what is known to be the dominating parts of NLL \cite{Salam:1999cn}.
\begin{itemize}
\item \emph{The running coupling} can be taken in account by replacing $\alpha_s$ with $\alpha_s(\mu)$ in the emission probability. The scale $\mu$ is set by the largest $p_T$ involved in the emission.
\item The \emph{non-singular terms} in the splitting function suppress large $z$-values, where the emitted parton takes the larger part of the momentum. Most of this effect is included by energy conservation, and requiring that the emitted partons are ordered in lightcone momentum $p_+$.
\item The \emph{energy scale terms} are essentially equivalent to projectile-target symmetry and are taken into account by not only requiring ordering in $p_+$, but also in $p_-$.
\end{itemize}
\item \emph{Confinement} is added by giving the gluon a mass. That modifies the splitting probability to have an exponential suppression for large transverse distances. This slows down the increase in cross section with energy.
\item \emph{Saturation} is present in the interaction in Muellers original formulation through multiple interactions. This only allows for loops that are cut by the interaction frame, not including loops contained in one of the cascades. To fully include saturation, and to restore frame independence, a 2 to 2 dipole ``swing'' is introduced. The swing can be interpreted either as a quadrupole correction, or as the exchange of virtual gluons. It tends to swing larger dipoles into smaller dipoles, which suppresses the growth of the cross section, and can give rise to 2 to 1 mergings in the final state, if one of the outgoing dipoles is reabsorbed.
\item \emph{Coherence} is important when a long distance emission is made from a parton with other partons close by. The long transverse wavelength of the emitted gluon cannot resolve the group of closely spaced partons, and the emission will treat the group of partons as one in terms of recoil and ordering. Without coherence, the restrictions of $p_-$ ordering would be overestimated, and the full phase space of allowed emissions would not be taken into account.
\end{itemize}

\section{The Interaction}
The elastic scattering amplitude between two colliding dipoles at a given impact parameter $b$ is in Muellers formulation given by
\begin{equation}
  f_{ij} = f(\pmb{x}_i,\pmb{y}_i|\pmb{x}_j,\pmb{y}_j) =
  \frac{\as^2}{8}\biggl[\log\biggl(\frac{(\pmb{x}_i-\pmb{y}_j)^2
    (\pmb{y}_i-\pmb{x}_j)^2}
  {(\pmb{x}_i-\pmb{x}_j)^2(\pmb{y}_i-\pmb{y}_j)^2}\biggr)\biggr]^2.
\label{eq:dipamp}
\end{equation}
This formula is in our model replaced by a similar calculation in $p$-space rather than $x$-space, and modified in the same way as the emission in the cascade for recoils and lightcone ordering, NLL effects, confinement and coherence. Saturation is already present in Muellers form through multiple interactions. The details of these effects are mainly guided by frame invariance, requiring that the corrections are the same in the interaction as in the evolution.

Summing over all pairs of dipoles, one from the state incoming from the negative $z$-axis, and one from the positive $z$-axis, gives the full first order elastic amplitude $F(\pmb{b})=\sum f_{ij}$. Allowing multiple interactions gives the unitarised elastic amplitude
\begin{equation}
T(\pmb{b})=1-e^{-F(\pmb{b})}.
\label{tf-relationmueller}
\end{equation}
Averaging over the incoming cascades, the interaction eigenstates, from both states now gives the elastic cross section
\begin{eqnarray}
d\sigma_{\text{el}}/d^2b &=& \langle \Psi_{\text{in}} | T | \Psi_{\text{in}} \rangle^2 = \langle \sum c_n \phi_n \Big| \, T \, \Big| \sum c_n \phi_n \rangle^2 \nonumber\\
 &=& \left( \sum c_n^2 T_n \right)^2 = \langle T\rangle^2
\end{eqnarray}
where $\Psi_{\text{in}}$ is the incoming mass eigenstates and $\phi_n$ are the interaction eigenstates, that is all pairs of dipole cascades, with weights $c_n$. $T_n = 1-e^{-\sum_{ij \in n} f_{ij}}$ is the scattering eigenvalue for the pair of cascades $n$. By replacing $\Psi_{\text{in}}$ with $\Psi_{X}$ the diffractive excitation cross section can be calculated, and the optical theorem gives the total cross section. From that, the non-diffractive cross section can be found:
\begin{eqnarray}
d\sigma_{\text{diff + el}}/d^2 b
&=&\sum_X \langle \Psi_{\text{in}} | T | \Psi_{X} \rangle \langle \Psi_{X} | T |
\Psi_{\text{in}} \rangle =\langle T^2 \rangle. \nonumber\\
d\sigma_{\text{tot}} / d^2b&=& 2\langle \Psi_{\text{in}} | T | \Psi_{\text{in}} \rangle = 2 \langle T \rangle \\
d\sigma_{\text{non-diff}}/d^2b& = & d\sigma_{\text{tot}} / d^2b-d\sigma_{\text{diff+el}} / d^2b  = \langle 1-e^{-2F}\rangle, \nonumber
\label{eq:eikonalcross}
\end{eqnarray}
where eq.~(\ref{tf-relationmueller}) was used in the last equality. Identifying $e^{-2F} = e^{-2\sum f_{ij}}$ as the non-interaction probability, the non-interaction probability of the individual dipole pairs factorises, and each pair of dipoles $i,j$ has a non-diffractive interaction probability of $1-e^{2f_{ij}}$. From this it is possible to select which dipoles interact in each collision.

\begin{figure}
\includegraphics[width=0.8\linewidth]{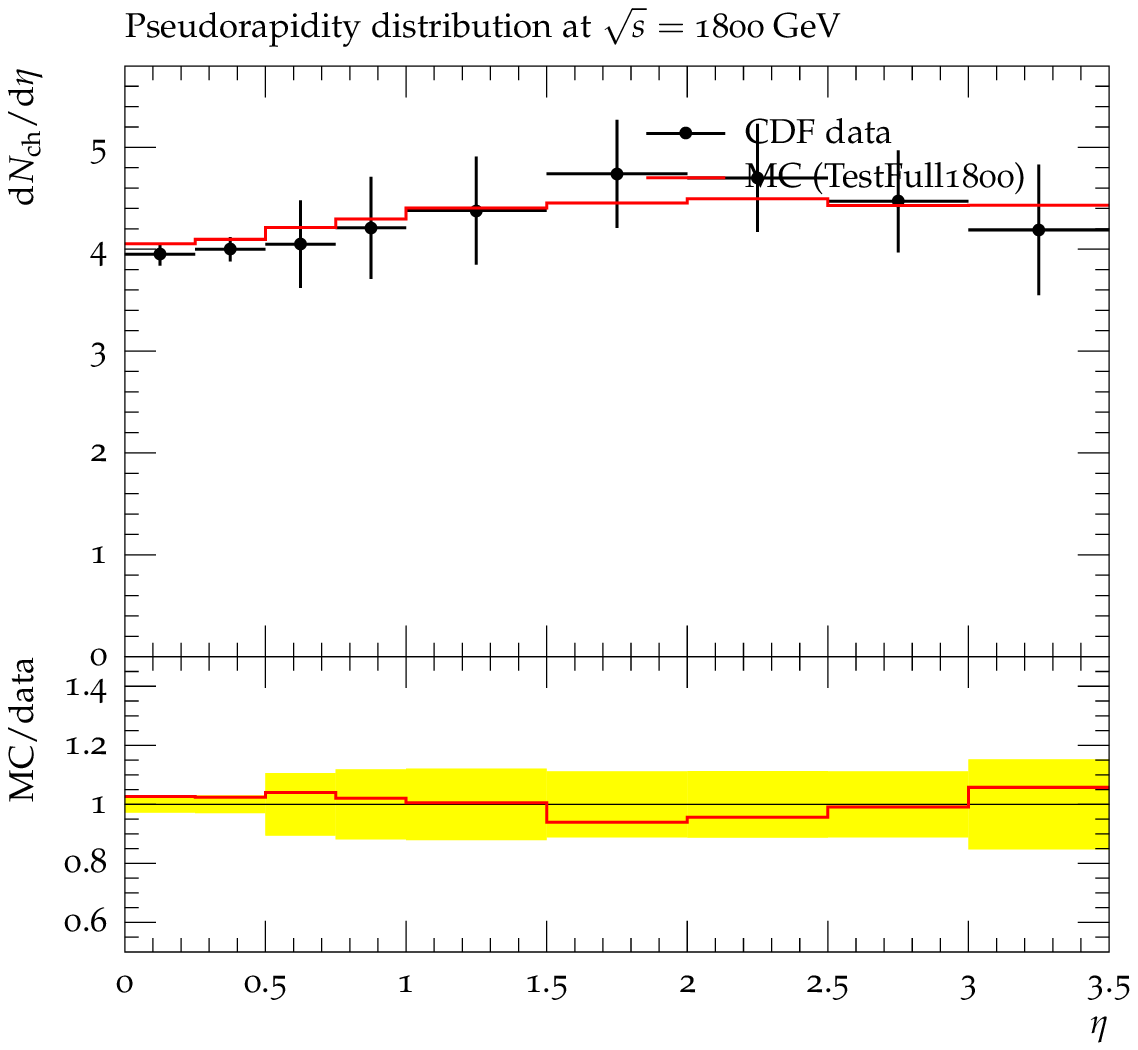}

\includegraphics[width=0.8\linewidth]{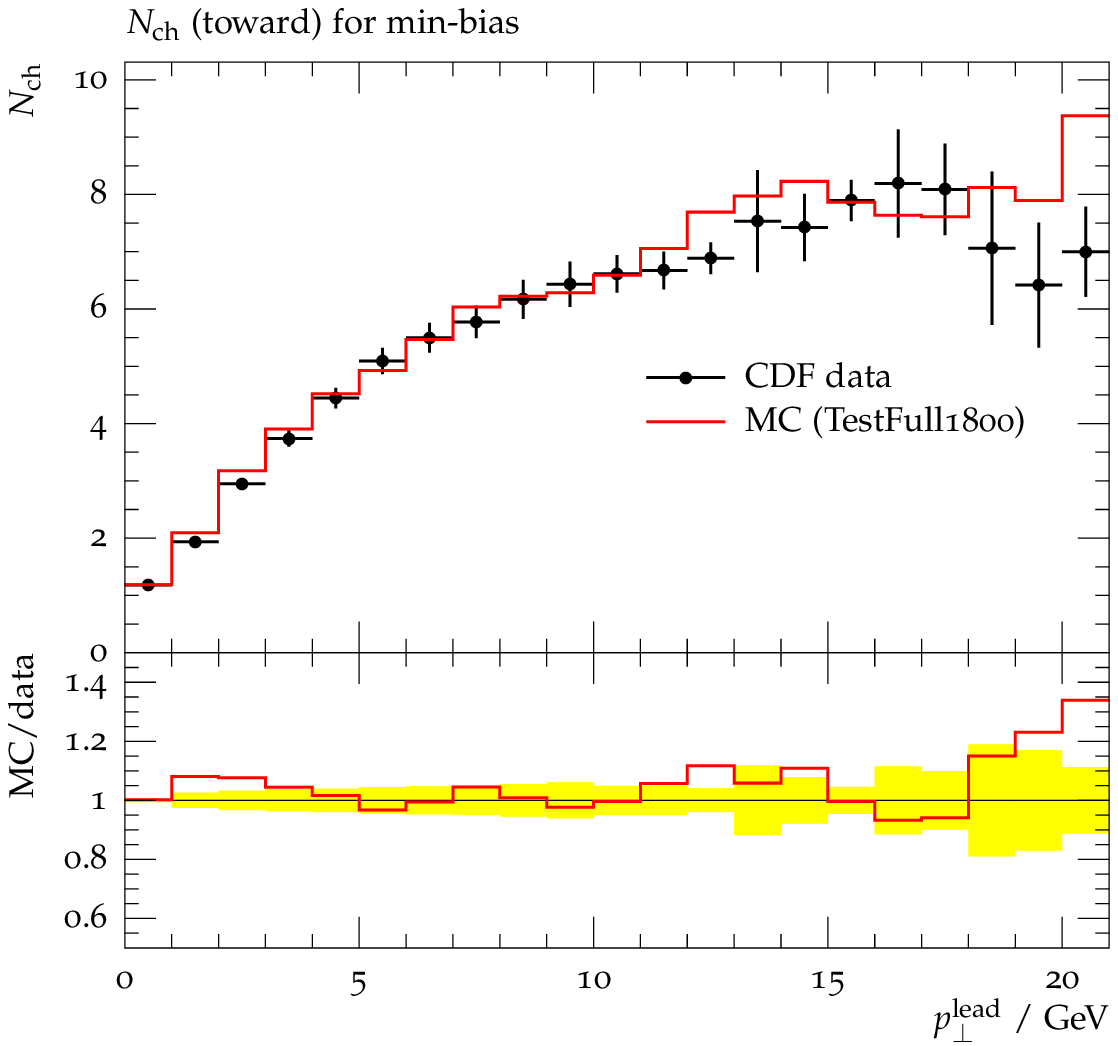}

\includegraphics[width=0.8\linewidth]{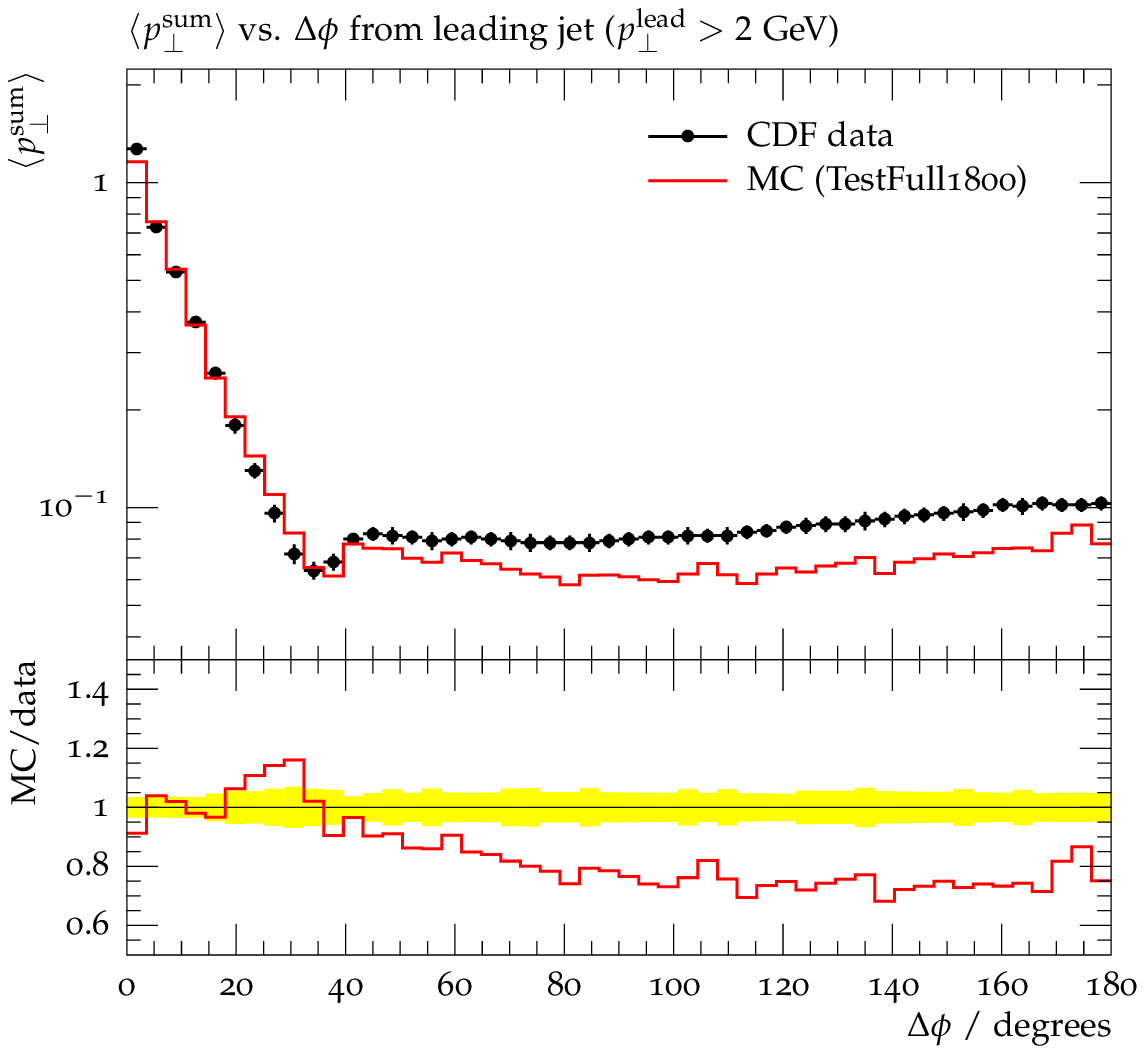}
\caption{ \label{fig:CDF} Comparison with a selection of data from CDF \cite{Abe:1989td,Affolder:2001xt}. The plots have been generated with Rivet \cite{Buckley:2010ar}.}
\end{figure}

\section{Finding the final state}
By tracing the interacting partons back towards the valence partons, the interacting gluon chains can be found. Since the timelike part of the shower will be regarded as final state radiation taken care of by ARIADNE, only the emissions directly connected to the interacting chains are kept. All other emissions are reabsorbed, to avoid double counting with the final state radiation.

\subsection{High $p_T$ suppression}
As can be seen from eq.~(\ref{eq:dipkernel1}) dipoles are in the cascade emitted with a weight of $d^2r/r^2$, corresponding to $d^2p_T/p_T^2$. However, the maximum $p_T$ in an interaction, coming from the cascade or from the interaction, should come with a weight $d^2p_T/p_T^4$. Muellers original model was designed for inclusive cross sections, which were not affected by small dipoles, and this discrepancy was not a problem. To produce a correct $p_T$-spectrum in the final state though, it is necessary to reweight the maxima. Reweighting according to \cite{Andersson:1995ju}, maxima in $p_T$ are reabsorbed with a probability
$$P_{\text{abs}} = 1 - \frac{p_{T\text{min}}^2}{p_{T\text{max}}^2},$$
where the maximum $p_T$ is compared to the closest minimum along the parton chain.

The extra phase space opened up by these reabsorbtions (mainly the $p_-$ ordering is significantly changed) is approximately covered by the coherence in the cascade: A maximum in $p_T$ means that the reabsorbed dipole is smaller than the proceeding emission, and thus that emission will not have resolved the small dipole, and the full phase space of the merged dipole has been taken into account.

\subsection{FSR and hadronisation}
The gluon chains connected by the exchanges in the interaction, together with the high-$p_T$ reabsorbtions above, determine the ISR. The partons and their colour structure are then passed on to ARIADNE \cite{Lonnblad:1992tz} to perform FSR. The phase space is restricted to spacelike emissions, matching the restriction in DIPSY.

After the FSR, the colour dipoles are allowed to fragment into hadrons as colour strings using PYTHIA \cite{Sjostrand:2006za,Andersson:1983ia}. Here the low $p_T$ cutoff in ARIADNE is matched to PYTHIAs starting scale to cover all phase space without double counting.

\begin{figure}[t]
\includegraphics[width=0.8\linewidth]{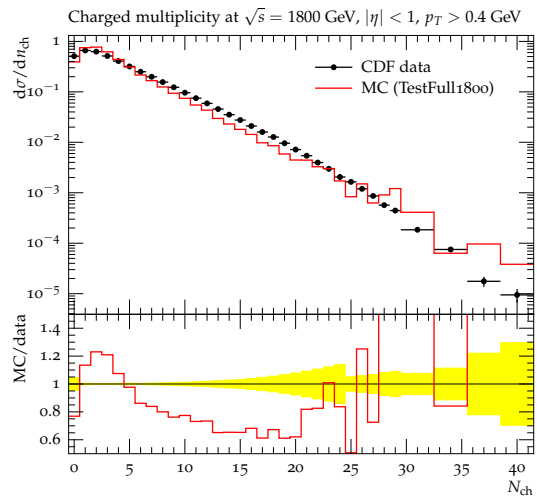}
\caption{ \label{fig:CDF2} Comparison with a selection of data from CDF \cite{Acosta:2001rm}. The plots have been generated with Rivet \cite{Buckley:2010ar}.}
\end{figure}

\section{Results}
As can be seen in previous publications, the following inclusive cross sections are well described at high energy:
\begin{itemize}
\item Total and elastic $pp$ as function of $\sqrt{s}$ and $t$,
\item Single and double diffractive excitation in $pp$ as function of $M_X$,
\item Total $\gamma^*p$ as function of $W$ and $Q^2$ (also low $Q^2$ with VMD),
\item Deeply virtual compton scattering as function of $W$, $Q^2$ and $t$ (also low $Q^2$ with VMD),
\item $\gamma^*p \rightarrow \rho p$ as function of $W$, $Q^2$ and $t$ (also low $Q^2$ with VMD).
\end{itemize}
The exclusive final state extension introduces many ambiguities from higher orders or non-perturbative QCD, which cannot easily be solved from first principles. All the plots are from a preliminary version of DIPSY from spring 2010. We will here discuss how it compares to data, and how sensitive the observables are to details in the algorithm.

\begin{figure}
\includegraphics[width=0.9\linewidth]{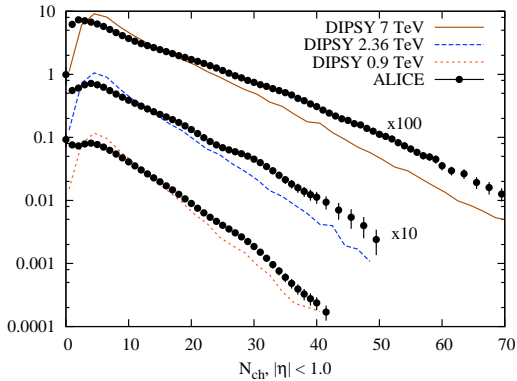}
\caption{ \label{fig:ALICE} Comparison with a selection of data from ALICE \cite{Aamodt:2010ft,Aamodt:2010pp}.}
\end{figure}

A selection of data is shown in figs \ref{fig:CDF}, \ref{fig:CDF2} and \ref{fig:ALICE}.
\begin{itemize}
\item The pseudorapidity distribution fits well with data in this preliminary version of DIPSY, but it is sensitive to details regarding mainly how the $p_\pm$ ordering is implemented.
\item The leading $p_T$ distribution in region close to the trigger particle fits well with data and is relatively stable between different versions. A successful prediction.
\item The distribution in $\Delta \phi$ around a minijet does not have enough activity in the transverse and away region. This is related to the rapidity distance between a minijet and its recoil. Possibly it is caused by an overestimate of the absorbtion of partons due to local $p_T$-maxima, giving too large rapidity distances between the remaining partons.
\item The multiplicity distribution from CDF follows data approximately over several orders of magnitude. This holds true for most version of DIPSY. The bias towards low multiplicities in this version is possibly related to the same overestimate of reabsorbtion. The same effect is seen in the ALICE multiplicity distributions, but at the higher energies DIPSY undershoots data somewhat in the tail.
\end{itemize}
The suppression algorithm for high $p_T$ for this preliminary version of DIPSY has a known flaw, allowing some small dipoles to survive unsupressed. This gives a too strong tail in the $p_T$ spectrum, but is corrected in later versions.

$\text{d}N_{ch}/\text{d}\eta$ in midrapidity at ALICE overshoots simulations at high energies in this version, but it is sensitive to several details in the model. More analysis is needed to understand the energy dependence in DIPSY.

\section{Conclusions and Outlook}
We have over the last years developed a BFKL based dipole model in transverse impact parameter space, including the major parts of NLL, confinement and saturation. This model has in previous publications shown to reproduce a wide range of total, elastic and diffractive cross sections with few parameters. Now we are producing exclusive final states with the same model. There are however many effects in higher orders, which it is not clear how to include, but that still has a significant effect on observables. Some effects can be accounted for with intuitive solutions, while others require tuning from data. The model can reproduce many exclusive observables at the Tevatron and the LHC. It should be noted that the results here are from a preliminary version of DIPSY, with several known flaws, and we expect the results in an upcoming publication to be improved.

While all data here are from $pp$ colliders, the model can be applied to any high energy collisions, as for example $DIS$, $AA$, $pA$ and $eA$. $AA$ will be the first to be studied after $pp$.

We would in the future also like to extend the model to include diffractive final states, but since the interaction amplitude does not factorise as neatly as in the non-diffractive case, it is more complicated. We do have some ideas though, and hope to return to this topic in future publications.

\bibliography{ismd08}

\end{document}